# High-precision molecular dynamics simulation of $UO_2$–$PuO_2$: Anion self-diffusion in $UO_2$

S.I. Potashnikov[a], A.S. Boyarchenkov[a], K.A. Nekrasov[a], A.Ya. Kupryazhkin[a]

[a] Ural Federal University, 620002, Mira street 19, Yekaterinburg, Russia

potashnikov@gmail.com  boyarchenkov@gmail.com  kirillnkr@mail.ru  kupr@dpt.ustu.ru

**Abstract**

Our series of articles is devoted to high-precision molecular dynamics simulation of mixed actinide-oxide (MOX) fuel in the approximation of rigid ions and pair interactions (RIPI) using high-performance graphics processors (GPU). In this article we study self-diffusion mechanisms of oxygen anions in uranium dioxide ($UO_2$) with the ten recent and widely used sets of interatomic pair potentials (SPP) under periodic (PBC) and isolated (IBC) boundary conditions. Wide range of measured diffusion coefficients (from $10^{-3}$ cm$^2$/s at melting point down to $10^{-12}$ cm$^2$/s at 1400 K) made possible a direct comparison (without extrapolation) of the simulation results with the experimental data, which have been known only at low temperatures (T < 1500 K). A highly detailed (with the temperature step of 1 K) calculation of the diffusion coefficient allowed us to plot temperature dependences of the diffusion activation energy and its derivative, both of which show a wide (~1000 K) superionic transition region confirming the broad λ-peaks of heat capacity obtained by us earlier. It is shown that regardless of SPP the anion self-diffusion in model crystals without surface or artificially embedded defects goes on via exchange mechanism, rather than interstitial or vacancy mechanisms suggested by the previous works. The activation energy of exchange diffusion turned out to coincide with the anti-Frenkel defect formation energy calculated by the lattice statics.

## 1. Introduction

In the previous article [1] we started comparison of the 10 sets of pair potentials (SPP) used for molecular dynamics (MD) simulations of uranium dioxide (three of these sets allow simulation of $PuO_2$ and mixed (U, Pu)$O_2$ fuel as well). Temperature dependences of lattice parameter, linear thermal expansion coefficient, bulk modulus, enthalpy, heat capacity (isobaric and isochoric), temperatures of melting and superionic transition were obtained and analyzed. In this article we will discuss the characteristics of mass transport, which plays a significant role in the manufacturing, exploitation and recycling of the nuclear fuel.

In particular, the diffusion of intrinsic ions determines such processes as formation and migration of defects and radioactive fission products (RFP), sintering, recrystallization and creep. As a result of these processes, the optimal operation mode is disrupted, the RFP segregation rate is changed, and the fuel degrades. Besides, recent studies show that burnup plays a major role in uranium dioxide ($UO_2$) transport properties and significantly changes thermal conductivity [2]. Nuclear fuel rods operate under strong temperature gradients (~$4 \times 10^5$ K/m), which causes changes in stoichiometry due to oxygen redistribution [3]. In turn, increase of stoichiometry lowers thermal conductivity, which increases temperature gradients even further. If a large amount of oxygen is redistributed, uniformity of mechanical properties might also be lost, which may reduce the mechanical stability of the pellet [4].

Therefore, in order to improve safety and minimize accident-related environmental damage of nuclear reactors, it is important to study the mass transport in $UO_2$ (which is currently the most used nuclear fuel [5]) over a wide range of temperatures from 700–1500 K in the operating mode up to 3150 K at the melting (due to Reactivity-Initiated Accidents, e.g. after the Loss Of Coolant).

Experimental data on $UO_2$ transport properties are known for temperatures below 1500 K [6] [7] [8] [9], which correspond to its crystalline state with relatively low concentration of lattice defects. These data do not directly describe processes of superionic transition, during which anionic defect concentration grows exponentially and then saturates at temperatures above 2600 K [10]. In particular, the exponential growth of creep rate near the superionic transition cannot be explained by extrapolating the low-temperature data [11].

Simulations by the method of molecular dynamics (MD) easily overcome difficulties of natural experiments [9], such as absence of suitable oxygen radioisotopes, control of stoichiometry and impurities, interaction between sample and containment, lack of thermostability of measurement tools. However, due to limitations of computational tools, MD simulations gave so far only a rough representation of intrinsic ions diffusion in $UO_2$. Some authors confined themselves to calculating the diffusion coefficient (DC) at a single temperature [12] [13]. Others plotted the temperature dependences of DC [14] [15] or mean square displacement (MSD) of ions [16] [17] with linear scale of both axes. Also, such representations of simulation results do not allow analyzing diffusion mechanisms, being suitable only for rough comparison of DCs in a narrow temperature range. Particularly, the change in the diffusion activation energy during superionic phase transition is difficult to detect when relying only on the charts with linear scale of DC, in which one can hardly distinguish two exponential functions with different activation energies. Consequently, the investigation of this transition on the basis of such charts led to ambiguous conclusions.

Analysis of DC in a wide temperature range, comparison with experimental data and calculation of diffusion activation energy require plotting a chart with the logarithmic scale of DC. Such charts for self-diffusion of oxygen anions in $UO_2$ are given in the following papers: [18] [19] [20] [21] [22] [23] [24] [25]. However, as it can be seen from Table 1, all the results obtained there (except [20]) have such limitations as a small number of temperature points (with a coarse step of 100–250 K) and narrow range of temperatures (only one or two points below 2200 K). These limitations affected the accuracy of determination of diffusion activation energy and made it almost impossible to investigate its temperature dependence in relation to the superionic transition. Such limitations persist even in 2012 when using a supercomputer as in the recent work of Devynck et al. [25].

Due to the limited amount of data, the authors of previous works were confined to process the dependences in Arrhenius coordinates (ln(D) vs. 1/kT) using a single straight line, which gives a single value of the diffusion activation energy for all the temperatures. Some authors fitted a line through all the points [18] [23] [25], while others only fitted a line through few low-temperature points obtained with the largest error [19] [21]. All of them had received the activation energy of 2.3–2.7 eV suggesting agreement with the experimental data. However, the degree of correspondence of the model to diffusion in the natural crystals remained unknown, because the simulated temperatures above 2000 K were too far from the temperatures studied experimentally (below 1500 K).

Besides, the results of previous works are contradictive. Thus, Arima et al. [24] admitted large activation energy (5.2–5.7 eV) and deviation of it from the low-temperature experimental data. Furthermore, in the papers [21] and [24] a deviation from linearity in the Arrhenius chart of DC dependence was noticed, whereas the authors of [18], [23] and [25] did not found it. However, Yakub et al. [21] claimed vacancy or interstitial diffusion mechanisms in crystalline phase and complex cluster mechanisms at high temperatures, Arima et al. [24] claimed "lattice diffusion" at moderate temperatures (but they did not specified, if it was interstitial or exchange mechanism) and vacancy diffusion at low temperatures in the case of artificial Schottky defects embedding. Govers et al. [23] stated domination of the exchange diffusion (four or eight ions at a time) in the whole temperature range. Therefore, the question of anion diffusion mechanisms in the model $UO_2$ crystals still remains open.

We also note that the results of the work [19] were obtained in NVT simulation (without a barostat) using the recommended lattice parameter [26], which exceeds the equilibrium one for the potentials Walker-81 used there by 0.2 Å (see measurements in a previous work [1]). Such a stretching of model crystal corresponds to a huge negative pressure (about −10 GPa), which leads to an increase in DC and a significant decrease in the superionic transition temperature due to a sharp acceleration in the processes of crystal lattice disordering.

Besides, our group has previously carried out a study of mass transport in $UO_2$ nanocrystals surrounded by vacuum (due to Isolated Boundary Conditions – IBC). In the paper [20] we measured DC of anions in the bulk and on the surface of nanocrystal using MOX-07 potentials and determined dominant diffusion mechanisms in contiguous temperature ranges. The anion diffusion activation energy in the bulk was found to depend on temperature, having the values of 2.5 eV, 3.2 eV and 1.3 eV in the ranges of 1400–1750 K, 1800–2600 K and 2650–3050 K correspondingly. The first two values were attributed to vacancy and interstitial diffusion under the anti-Frenkel disordering of the crystalline phase, and the latter was ascribed to the direct interstitial migration in the superionic phase. It is important to note, that formation of the separate 1400–1750 K segment was essential to provide an agreement of the model with the experimental data. However, the coarse temperature step of 50 K obstructed analysis of the superionic transition and the effects of nanocrystal structural relaxation discussed in [27] were not taken into account. Moreover, the influence of potentials and boundary conditions upon the model was not studied. Later the anion self-diffusion in nanocrystals was also studied with Walker-81 and Nekrasov-08 SPPs [22], but that work lacked segmentation of diffusion mechanisms and the aforementioned issues were not overcome.

Finally, the approximation of rigid ions and pair interactions (RIPI) under periodic boundary conditions (PBC), which was used in most of the previous works, could affect the results of diffusion simulation. Particularly, existence of the superionic transition could be affected due to either absence of polaron and Schottky defects or inevitable movement correlation of all periodic defect "reflections". Consequently, it is necessary to investigate the scope of applicability of aforementioned approximations as well as dependence of the simulation results on SPP and system size (i.e., ion count).

## *2. Methodology*

As in our previous works [1] [28], molecular dynamics simulations were carried out in the rigid ions approximation for the ten sets of empirical pair potentials: Walker-81 [29], Busker-02 [30], Nekrasov-08 [22], Goel-08 [31], obtained in the harmonic approximation from elastic properties at zero temperature; Morelon-03 [18], obtained using the lattice statics from energy of Frenkel and Schottky defects; Yamada-00 [16], Basak-03 [17], Arima-05 [32], MOX-07 [33] and Yakub-09 [34], obtained using MD simulation of thermal expansion and bulk modulus. Parameters of all these SPPs are given in Tables 1–3 of the first article [1].

Similarly, in this work we use the technology of parallel computing on high-performance graphics processors [20] [35] [36]. This approach allowed us to obtain the diffusion coefficients of oxygen anions in a wide range of $10^{-11}$–$10^{-3}$ cm$^2$/s (not attained before) and temperatures from 1400 K up to the melting points of model crystals, thus providing the possibility of direct comparison of simulation results to the low-temperature experimental data without recourse to extrapolations. The temperature dependences of the anion self-diffusion coefficient $D(T)$ are calculated using a fine temperature step of 1–5 K (increasing the number of simulations up to several thousand for each SPP), which is necessary to follow up the temperature dependence of diffusion activation energy (and its derivatives) and to detect any changes in the migration mechanism.

First of all, in order to calculate the point defects formation energies by the method of lattice statics, a gradient optimization of ion coordinates under PBC was used, but, in contrast to the Mott-Littleton approach (see, e.g., [37]), relaxation was carried out throughout the whole crystal lattice (instead of just inner sphere) so that the calculated values correspond to intrinsic defects which could arise during the MD simulation.

The defects studied by the lattice statics were the Frenkel pair (FD) consisting of interstitial cation and cationic vacancy, the anti-Frenkel pair (AFD) consisting of interstitial anion and anionic vacancy and the Schottky defect (SD) consisting of electroneutral combination of one cationic and two anionic vacancies (i.e., trivacancy). Each energy value was calculated in stoichiometric and electroneutral crystal (instead of sum of energies of the crystal with vacancy and the crystal with interstitial ion as in [37]), where the interstitials and the vacancies were formed at maximum distance possible within periodically-translated supercell.

In order to eliminate the influence of defect "reflections" appeared under PBC, we plotted the dependences of the defect formation energy on the supercell size, which ranged from 4×4×4 up to 16×16×16 unit cells of 12 ions each (the number of unit cells per edge of supercell is denoted by C below). These dependences were found to be linear with respect to 1/C (see Fig. 1), which allowed us to calculate the defect formation energies at infinite distance (between the vacancy and the interstitial ion in case of either FD or AFD and between the vacancies in case of SD). The similar methodology of defect energy calculation was used in the recent work of Devynck et al. [25]. However they did not excluded energies for the smallest supercell (2×2×2 and 3×3×3) and had to use a cubic fit, which could slightly lower the extrapolation results.

The results of such extrapolations, together with the relevant data obtained by Mott-Littleton technique (from [37] and [38]) are shown in Table 2 and compared with ones obtained by other authors using the density functional theory (DFT) and with the diffusion activation energies ($E_D$) obtained from our MD simulations.

All MD simulations in this work were carried out on high-performance graphics processors (GPU) using NVIDIA CUDA and Microsoft DirectCompute technologies, which gave us speedup of 100–1000 times (see details in [35] [36] [41]). We used Ewald summation of long-range ionic interactions. In order to integrate Newton's equations of motion, the Verlet method (with time step of 5 fs) and Berendsen thermostat-barostat (with a relaxation time of 1 ps) were used. The first 5 ps (1000 MD steps) of NPT-simulation (with a constant number of particles, pressure and temperature) were spent to reach equilibrium.

The diffusion coefficient D of intrinsic ions of type $S$ at a given temperature $T$ can be calculated from linear time ($t$) dependence of the mean square displacement (MSD) of ions by the Einstein relation:

$$\lim_{t \to \infty} \langle \Delta R^2 \rangle_S = 6tD_S(T) + const$$

$$\langle \Delta R^2 \rangle_S = \frac{1}{N_S} \sum_{i \in S} |R_i^t - R_i^0|^2 \quad (1)$$

Here, at the time moment $t$ the squared displacements of the particles with respect to their initial positions are summed up over all particles of type $S$. These initial positions were updated every few nanoseconds in order to eliminate systematic errors, caused by arbitrary choice of starting moment of time and computational errors accumulation as well.

On the assumption of a persistent diffusion mechanism, the temperature dependence of the diffusion coefficient is expressed by the Arrhenius equation $D(T) = D_0 \exp(-E_D / kT)$, where the diffusion activation energy $E_D$ and the coefficient $D_0$ are constant characteristics of the process. In practice this equation holds approximately for the temperature intervals where a distinguished mechanism of migration exists. Such intervals are shown in Tables 3–5.

The rate of diffusion jumps decreases exponentially with decreasing temperature. This fact requires exponential increase of either simulation time or supercell size in order to preserve statistically acceptable accuracy of DC measurement via MSD. However, in many previous studies the authors tended to increase an ion count (up to 20 736 in [23]), rather than a simulation time. In the case of too rare jumps linear MSD curves become stepped (see Fig. 2), and in the absence of diffusion jumps the MSD values cannot be used to obtain the diffusion coefficient, reflecting the thermal oscillations only. For example, if the model UO$_2$ crystal consists of 1500 ions, then a simulation during 0.1 ns would allow obtaining DC down to values of ~10$^{-7}$ cm$^2$/s only.

In order to catch optimal number of diffusion jumps, that is sufficiently large to achieve a reasonable error and sufficiently low to avoid redundant computations, we continued each computing experiment until the threshold MSD value of 10 Å$^2$ was reached (during not less than 0.01 ns at high temperatures and up to 1000 ns at low temperatures where DC reaches ~10$^{-11}$ cm$^2$/s). For example, MSD value of 10 Å$^2$ corresponded to the simulation time of 0.01, 0.1 and 1 ns at respective temperatures of 3100 K, 2460 K and 2265 K (Fig. 3).

It also can be seen in Fig. 3, that in the case of diffusion jumps deficit, the diffusion coefficient is often decreased stronger then it is increased due to logarithmic scale. Exclusion or recalculation of these decreased values (which are obviously invalid) instead of additional accumulation of statistics leads to overestimation of DC and underestimation of diffusion activation energy. Binding of simulation time with accumulation of some minimal MSD value, which is suggested above, eliminates this systematic error. As a result, in figures of this work the uncertainties are smaller than the symbol size of curves.

Besides, low number of points with high uncertainty also complicates processing of data. We believe that this happened in the work [21], where the value of E$_D$ = 2.3 eV was calculated from the three lowest temperature points, probably having the largest statistical error. However, if the authors had not excluded their points from the range of 2300–2750 K, they could receive E$_D$ = 5.2 eV, which is much closer to the AFD formation energy calculated by the lattice statics and to our value of 5.7 eV (obtained with high statistical accuracy due to fine temperature step and guaranteed number of diffusion jumps). Unfortunately, they believed that E$_D$ exceeding the recommended experimental value of 2.6 eV had no direct physical meaning.

The influence of system size was studied by variation of ion count from 324 up to 49 152 particles. Similar to the simulation of thermophysical properties (see [1]), the behavior of the system of 1500 particles was quantitatively almost the same as behavior of larger systems (see Fig. 4), and for half of the ten SPPs considered the differences between the systems of 768 and 1500 particles were quite small too.

So it is profitable to increase simulation time, not system size, as the workload depends linearly on the number of MD steps, but quadratically on the number of particles. In this work a quite long simulation time of 1000 ns was reached, which is statistically equivalent to 10–40 ps simulation (used in previous works, as seen in Table 1) of 40–160 million ions, but is by 4 orders faster due to O(N$^2$) computational complexity. Even if there was a quick O(N) algorithm, the system size still could not be used for dynamical control of count of diffusion jumps, because it should be chosen at the beginning of simulation.

In addition to the periodically-translated crystals without surface, we also simulated the nanocrystals isolated in vacuum, which had free surface of energetically-optimal octahedral shape as discussed in [27], in order to study the influence of boundary conditions. The volume of such system was not fixed, so nanocrystal simulations were carried out under NPT ensemble with zero ambient pressure.

The mobility of ions in the bulk and at the surface of nanocrystal differed by up to 5 orders of magnitude [20], therefore these two regions should be considered separately. In this work we were interested only in bulk diffusion, because it can be directly compared with diffusion simulated under PBC and with experimental results. The MSD was summed up over anions within such sphere that excluded all surface particles. Ions that moved from the bulk to the surface were also excluded. Since the number of such excluded ions is increasing with time, the initial positions of particles should be updated regularly.

For each SPP a series of simulations was carried out with temperature step of 1 K for D(T) > 10$^{-9}$ cm$^2$/s and with step of 5 K for $D(T)$ < 10$^{-9}$ cm$^2$/s (advantages of such small temperature steps were justified in our previous work [1]). The curves of temperature dependences D(T) in Figs. 4–7 were plotted with a step of 100 K and averaging over an interval of ±100 K for smoothness. In addition to main results, Figs. 5 and 7 also show data for a molten UO$_2$ system "MOX-07 melt".

## 3. Results and discussion

### 3.1. Lattice statics calculation of defect formation energies

The results of lattice statics and *ab initio* DFT calculations in comparison with MD simulation results and experimental data are shown in Table 2.

It is known [8] [9] [42], that in stoichiometric uranium dioxide, as well as in other crystals with the fluorite structure, the dominant type of anionic sublattice disordering is anti-Frenkel defect with formation of interstitial anions and anionic vacancies. Vacancy concentration can be directly measured in a wide temperature range by technique of high-temperature diffraction and inelastic neutron scattering (see, e.g., [43]), which allows to determine the energy of AFD formation. Note that interpretation of available data remains ambiguous due to the complexity of measurement conduction and analysis, as well as significant statistical errors. This is evidenced, in particular, by the difference of value (4.6 ± 0.5) eV from the work [10] and the Matzke's recommendation of (3.5 ± 0.5) eV derived from a large set of experimental data [8]. The most recent value of (3.8 ± 0.6) eV has been obtained from the coherent diffuse neutron scattering experiments of Clausen et al. [44] and can be treated as override of their previous work [43].

As it can be seen from Table 2, the AFD formation energy is reproduced well only by two SPPs, namely MOX-07 and Morelon-03. For both SPPs, the values are close to the Matzke's recommendation of $(3.5 \pm 0.5)$ eV, more recent experimental value of $(3.8 \pm 0.6)$ eV and the results of modern DFT-calculations ~4 eV (see the review [39]). However, Morelon-03 potentials, unlike MOX-07, were fitted to the defects formation energies proposed by Matzke [8], therefore the results for this SPP cannot be considered as independent estimates. The values of 5.2–5.8 eV obtained for the other potentials (Goel-08, Yakub-09, Nekrasov-08, Yamada-00, Basak-03 and Walker-81) exceed even the upper bound of Hutchings's estimate [10], and the highest values ~8 eV were obtained for Arima-05 and Busker-02.

The lowest values for energy of cationic FD formation belong to Morelon-03 and MOX-07 SPPs (15.6 eV for both), slightly larger values – to Yakub-09 (15.9 eV) and Basak-03 (16.8 eV). The calculated values of SD formation energy are also overstated, but accidentally the lowest values of 7.7–7.8 eV correspond to Walker-81 and Goel-08 SPPs, which gave quite large deviation of thermophysical properties from the experiment (see [1]). All these values are higher than Matzke's recommendations (see Table 2); however, there exist DFT calculations, which provide higher cationic defect formation energies, which are close to our results for MOX-07 and Yakub-09 SPPs: ~15 eV for FD and ~10 eV for SD [40].

The defect formation energies obtained in the rigid ions approximation for recent shell-core potentials of Read and Jackson [38] are 1.5 times higher than in the original work and almost as high as energies of Busker-02 and Arima-05, which support our hypothesis previously stated in [1] that shell-core potentials with formal charges are unsuitable for simulations without shells.

### 3.2. MD simulation of anion self-diffusion

In this study we calculated the diffusion coefficient (DC) of anions over wide range of temperatures from 1400 K up to the corresponding melting points (see [1]) with fine step of 1–5 K, which allowed $E_D(T)$ dependence investigation. Fig. 5 shows that for all SPPs considered the temperature dependences of DC have a form of hyperbola with two asymptotes: low-temperature line with low DCs and high effective diffusion activation energy and high-temperature line with high DCs and low activation energy. Hence we attribute these low-temperature and high-temperature regions to the crystalline and the superionic phase, respectively.

In addition, Fig. 5 shows that the transition between crystalline and superionic phases (i.e., anionic sublattice disordering) occurs gradually over a wide temperature range, which is related to the λ-peak region of heat capacity. Consequently, in determining the diffusion activation energy of asymptotes this transition region must be excluded.

Since the activation energy $E_D$ is dependent on temperature, it is important to investigate this dependence in order to get precise and reliable $E_D$ values for both phases and also to characterize the superionic transition. Fig. 7 shows $E_D(T)$ dependences obtained as derivatives of the Arrhenius dependences $\ln(D)$ vs. $(1/kT)$, which were double averaged over an intervals of ±100 K for smoothness. All of them, except discontinuous curve "Yamada-00", are S-shaped with a clearly distinguished flat plateaus corresponding to the crystalline and superionic phases, which are separated by transition region with a width of about 1000 K. Therefore, it is correct to determine a characteristic activation energy for each phase according to the temperature intervals of the plateaus. As the averaging slightly changed the length of plateaus, we took non-averaged data and adjusted these temperature intervals minimizing the standard deviation. Tables 3–5 show the resulting $E_D$ values and boundaries of the adjusted intervals. In order to determine the mechanisms of self-diffusion we used these values of the activation energy in conjunction with visual observation of the ions movement.

The visual observation through low-temperature range (see boundaries in Table 3) showed the absence of long-lived anti-Frenkel pairs even in a large system of ~50 000 ions (16×16×16 unit cells) simulated under PBC. In other words, there were no vacancies moving away from interstitial ions. Instead, every time ions were permutated cyclically with simultaneous formation of one or more adjacent anti-Frenkel pairs, which recombined shortly. This diffusion mechanism is often called exchange, although some authors distinguish direct exchange of two ions and ring substitution of more than two ions. Exchange diffusion probably exists in natural crystals, which can be supported by experimental observation of short-time anti-Frenkel pair formation in fluorite-type crystals of $SrCl_2$ (see the review [9]).

We found that exchanges usually start with the motion of two ions, when the first ion acquires additional energy and pushes its neighbor into the interstitial site, because this neighbor begins to move soon after the first ion. If this interstitial site is near the vacancy of the first ion, then recombination of the Frenkel pair occurs immediately (exchange of two ions). Otherwise, ions in turns occupy the existing vacancy, which is similar to the opposite motion of this vacancy. The closeness of interstitial ion and vacancy is energetically more efficient, hence recombination of the Frenkel pair is early and inevitable. The instability of configurations with interstitial ion was accompanied with higher amplitude of oscillations of neighbor ions, which sometimes assists concurrent exchanges in other parts of the crystal.

The simplest example of such exchange is illustrated by trajectories of two anions in Fig. 8 and their velocities in Fig. 9 (which were calculated from averaged ion positions in order to mitigate thermal oscillations and emphasize motion between lattice sites). The first ion starts to move with sharp acceleration towards the neighbor lattice site (see the steep slope in Fig. 9). The second ion begins to move after 7–10 MD steps. After 40–50 steps the second ion reaches an interstitial site (see the Fig. 8), while the first ion completes the movement to the vacant lattice site. Then the second ion accelerates again and during the following 30 steps occupies the lattice site, which was abandoned by the first ion. This process lasted ~0.4 ps, while the observed exchanges of four and seven ions lasted 1–1.3 ps.

We suggest the following explanation of the S-shaped dependence $E_D(T)$. Diffusion jumps are rare in the crystalline phase, so the emerging defects do not affect each other. That is why activation energy remains constant at sufficiently low temperatures (see the left plateaus in Fig. 7). With increasing temperature the jumps are more likely to occur in the presence of other defects, and interaction with them lowers the diffusion activation energy. It is possible to quantify somehow this interaction using Fig. 1, where the system size determines the distance between an anti-Frenkel defect and its periodic reflections. At sufficiently high temperatures near the right plateau of superionic phase the anions are less likely to encounter barriers in their way as the neighbors are constantly moving between lattice sites, so the activation energy ceases to decrease and gradually reaches a constant value.

As a result of complete disorder of anionic sublattice the difference between lattice site and interstice disappears, so the diffusion occurs without formation of intrinsic defects. The visual observation in this phase showed us that anion migration resemble free movement in a system of canals formed by the cationic sublattice. Correspondingly, the diffusion activation energy is reduced to the energy of interstitial migration (see Table 4).

Similar behavior is obtained in the neutron scattering experiments [45] [44], where the exponential increase of anti-Frenkel pair concentration changed into plateau at a temperature of 2600–2700 K. In our model as in those experiments the concentration of pairs during superionic transition reaches about 20% of the anionic sublattice sites, so the defects form inseparable system of vacancies and interstitial anions.

Thus, molecular dynamics simulation of $UO_2$ under PBC allows only three mechanisms of oxygen self-diffusion – one for each phase state (the ranges of activation energies are taken from Tables 3–5, where the lowest values correspond to MOX-07 SPP and the highest – to Busker-02 SPP):

- Molten state – free movement of anions with a minimal energy of 0.7–1.4 eV.
- Superionic state – indirect interstitial diffusion (extrusion of neighbors) without formation of intrinsic defects through "canals" in cationic (uranium) sublattice with energy of 1–2.5 eV, which is close to the energy of interstitial migration obtained by the lattice statics [18] [20] [37].
- Crystalline state – exchange diffusion, i.e. cyclic permutations of ions with formation of short-lived anti-Frenkel pairs, characterized by energy of 3.9–7.9 eV, which is close to the AFD formation energy obtained by the lattice statics.

It is interesting to note that behavior of cations just below the melting point is the same as behavior of anions below the superionic transition: they migrate via exchange mechanism characterized by diffusion activation energy close to the energy of FD formation obtained by the lattice statics (the details will be given in a future article).

All three mechanisms of anion self-diffusion were confirmed by visual observation of ions movement during MD simulations and are analyzed quantitatively in the following sections.

### 3.2.1. Low-temperature region

The lowest diffusion coefficients obtained in this work reached $4 \times 10^{-12}$ cm$^2$/s. Attainment of such low values required up to 1000 ns (200 million MD steps) of simulation time and made possible a direct comparison with the known low-temperature (T < 1500 K) experimental data (i.e., without extrapolations, which were necessary in the previous works).

Experimental measurements of oxygen diffusion in $UO_2$ are hampered as compared with other materials [8] [9]: firstly, due to inapplicability of direct tracer techniques caused by absence of radioactive tracer for oxygen; secondly, due to difficulty of exactly defining the stoichiometric composition (in particular, if powders are used). A variety of indirect methods using stable isotopes had been developed, but none of those methods could compare with the high precision of the direct techniques. As a result, there is a certain scatter in the available data.

The data of Marin and Contamin [6] with effective activation energy of 2.56 eV are generally accepted as being the most reliable [9], and the Matzke's recommended value of 2.6 eV is close to it. Matzke also suggested that it includes contributions of both vacancies and interstitial ions.

Our values of the low-temperature diffusion activation energy of 3.9–7.9 eV differ from the lower values of 2.3–2.6 eV obtained in the earlier studies [18] [19] [21] [23] for the potentials Walker-81, Basak-03, Yakub-09. However, their results are in contradiction with the corresponding values of AFD formation energy, which are rather high (5.6–6.0 eV, as seen in Table 2), as in ideal lattice the diffusion cannot occur without defect formation. Those activation energies could be underestimated due to small number of simulations and the subsequent misinterpretation of data. Particularly, in the works [18] and [23] a single straight line (lnD vs. 1/T) was used to process both superionic and crystalline phases which, therefore, were not distinguished from each other. As it was discussed in Section 2, the authors of [21] had excluded the points in the range 2300–2750 K in favor of the three lowest temperature points, which decreased the obtained activation energy nearly two-fold. Finally, as mentioned above, in [19] the measurements were performed under a large effective negative pressure of –10 GPa, which could result in a significant decrease of both superionic transition temperature and the diffusion activation energy.

Fig. 5 shows that only two of the potential sets (namely Morelon-03 and MOX-07) provided values of the oxygen diffusion coefficient which are close to the experimental data, while the other SPPs underestimated the DC by at least 2 orders of magnitude. However, even the lowest value of the effective diffusion activation energy (see Table 3) exceeded the Matzke's recommendation $E_D = 2.6$ eV by more than 1 eV.

Since the AFD formation energy of Morelon-03 and MOX-07 is close to the experimental estimates (see section 3.1 and Table 2), we can assume that the main reason for deviation of activation energy is difference of diffusion mechanisms. For all ten SPPs the AFD formation energy is close to $E_D$, which indicates that diffusion jumps comprise of AFD formation but not migration of the corresponding vacancies and interstitial ions. According to this, we visually observed exchange diffusion via ion permutations, as discussed in Section 3.2.

The defect formation energies provided in Table 2 correspond to infinitely separated Frenkel pairs. However, one can expect that $E_D$ values obtained from MD simulations should be compared with energy of bound Frenkel pairs. In attempt to calculate these energies using the lattice statics method we found that:
1) The nearest pair (with distance of $\sqrt{3}/4 \approx 0.43$ lattice periods) recombines immediately.
2) The second nearest pair (with distance of $\sqrt{11}/4 \approx 0.83$ lattice periods) turned out to be stable only for a few SPPs, which made impossible the comprehensive comparison. For MOX-07 SPP its energy is in the range of 2.8–2.9 eV.
3) The third nearest pair (with distance of $\sqrt{19}/4 \approx 1.09$ lattice periods) turned out to be stable. The energy of such pair highly correlates with energy of infinitely separated pair ($E_{1.09} \approx 0.83\ E_{inf}$), but correlation with $E_D$ (even with addition of estimates of vacancy migration energy ~0.2–0.5 eV) did not improved.

These results agree with the recent work of Devynck et al. [Devynck12] and support our belief, that static calculations of equilibrium configurations are not able to precisely predict energy of non-equilibrium processes such as dynamically simulated exchange diffusion, but can be used for qualitative estimates.

The fact that interstitial or vacancy diffusion mechanisms do not occur in such conditions regardless of system size and SPP leads to the conclusion that defect-free crystals simulated under PBC are not suitable for simulation of self-diffusion processes at low temperatures (below superionic transition).

However, we observed long-lived anti-Frenkel pairs in nanocrystals simulated under IBC (i.e., isolated in vacuum), which have free surface. With MOX-07 potentials the dominant exchange mechanism of oxygen migration was replaced by interstitial and vacancy migration under the "classical" anti-Frenkel disordering (i.e., long-lived pairs) at the temperatures below 1800 K [20]. Bend in diffusion coefficient dependence due to this changeover of mechanisms can be seen on the curve "MOX-07 IBC cube" in Fig. 6. Similar behavior was obtained in experiments on oxygen self-diffusion measurement in $ThO_2$ [46], which also has the fluorite lattice structure. However, cubic nanocrystals simulated in our previous work [20] are subject to structural relaxation (see [27]), so in order to exclude its effect on diffusion in this work we also simulated nanocrystals of the energetically-optimal octahedral form. As seen in Fig. 6, the curve "MOX-07 IBC octa" shows good agreement with the available experimental data being somewhat lower than "MOX-07 IBC cube". Its effective activation energy below 1800 K decreases down to $(2.9 \pm 0.3)$ eV, which is 1 eV lower than the value obtained under PBC.

In order to verify that such changeover of diffusion mechanism is due to surface and not boundary conditions, we simulated a crystal with embedded cavity (i.e., inner surface) under PBC. In Figs. 6–7 the curves "PBC 4 SD" correspond to simulation of 5×5×5 supercell with the cavity of one unit cell (which gives 4 trivacancies bound together, i.e. 0.8% concentration of Schottky defects). These curves show qualitatively the same behavior as isolated nanocrystal under IBC. In both cases there is a decrease in the diffusion activation energy below 2000 K, corresponding to a change of the dominant diffusion mechanism, while in the absence of surface no such a change is observed. Quantitatively, however, diffusion coefficients in a system with cavity are overestimated by two orders at 1500 K, and diffusion activation energy almost reaches 1 eV, because anions migrate via embedded vacancies without formation of additional anti-Frenkel pairs.

Thus, MD simulations under PBC (either with or without cavity) give diffusion mechanisms that are different from the mechanisms in natural crystals at low temperatures. Therefore, octahedral nanocrystals under IBC seem to be the best way to simulate diffusion in crystalline phase of uranium dioxide.

### 3.2.2. Superionic transition

As it was shown above, the model crystals without surface (or cavities) do not seem to allow correct simulation of either interstitial or vacancy diffusion mechanism at low temperatures. However, in the superionic transition region (above ~2000 K for MOX-07 SPP) both periodic and isolated boundary conditions give us almost the same dependences of oxygen diffusion characteristics (see Figs. 6–7), as the exchange mechanism via ion permutations is dominant in both models. The existence of this mechanism in natural $UO_2$ crystals at the temperatures above 2000 K is indirectly confirmed by neutron scattering experiments [45] [44], where high concentrations of the close anti-Frenkel pairs (in a form of small defect clusters) were observed. Thus, we consider it plausible to use PBC for MD simulations at the temperatures of superionic transition region up to the melting point.

The curves in Figs. 5–7 clearly indicate that model $UO_2$ crystals undergo a gradual disordering of anionic sublattice, which is manifested as fundamental decrease of the effective activation energy of oxygen self-diffusion. This behavior is observed with all the potential sets at sufficiently high temperatures except for the Yamada-00 SPP.

According to our previous work [1], Yamada-00 potentials give, instead of the gradual superionic transition, a zone of instability (with oscillations between crystalline and disordered states) and a first order phase transition featured by abrupt change in lattice parameter and enthalpy. As seen from the subfigure in Fig. 5, in this work the diffusion coefficient and the activation energy also change abruptly within a region of instability, while the superionic phase extends to unusually low temperatures ~2200 K (i.e. almost 500 K lower than current recommendation of 2670 K [26]).

The obtaining of wide intermediate region of the gradual superionic transition is an interesting result itself. Currently, there exist phenomenological hypotheses [26] [21] [47] of discontinuous transition between the two (crystalline and disordered) equilibrium states of anionic sublattice and the corresponding "narrow" (30–50 K) λ-peak of heat capacity with "infinite" height (like the Dirac delta function), from which it follows that the chart of oxygen diffusion coefficient (DC) in the Arrhenius coordinates should have a break, corresponding to a discontinuous and stepwise temperature dependence of diffusion activation energy (like the Heaviside step function). Those hypotheses influenced the interpretation of MD simulation results, e.g. in the article of Yakub et al. [21] the fact that diffusion data indicated a broad transition interval of temperatures was discarded (see discussion in Section 2). On the contrary, our precise calculations of both heat capacity (see [1]) and DC dependences with a step of 1 K are consistent with each other and indicate a wide and smooth phase transition (with a range of temperatures over 1000 K and λ-peak height below 0.03 kJ/(mol*K)) regardless of SPP and system size. As discussed in Section 3.2, the DC curves have a form of hyperbola with two asymptotes for crystalline and superionic phases, which corresponds to S-shaped curve of diffusion activation energy with two plateaus (for completely ordered and completely disordered anionic sublattice) and wide transition region between them (see Fig. 5–7).

Moreover, in [1] we obtained the dependence of specific heat with a characteristic λ-peak by differentiating enthalpy with respect to temperature. In this work, differentiation of temperature dependences of the diffusion activation energy also gives λ-shaped curves (see Fig. 10). They were averaged once more over intervals of ±100 K for smoothness. The width of the peaks is close to 1000 K, and the temperatures of their maxima are almost equal to the λ-peak temperatures of the isochoric heat capacity from [1]. Also, it can be noted that the curve $dE_D(T)/dT$ corresponding to Morelon-03 SPP has the lowest peak and nearly flat plateau, which means almost linear change in $E_D(T)$ during superionic transition. This behavior is possibly due to the complex spline shape of anion-anion interactions of Morelon-03.

### 3.2.3. High-temperature region

It is interesting that only the periodic boundary conditions allows studying of the high-temperature superionic phase per se, because under isolated boundary conditions the defect concentration does not reach saturation (see the curve for nanocrystals in Fig. 7) due to lower melting point (see [28]).

As discussed above, the main mechanism of the anionic sublattice disordering of the model $UO_2$ crystals is formation of the short-lived anti-Frenkel pairs. During superionic transition the concentration of the pairs reaches about 20% of the sublattice sites as in the experiments [45], so the defects form inseparable system of vacancies and interstitial anions. The visual observation in this phase showed us that anion migration resemble free movement in a system of canals formed by the cationic sublattice. Correspondingly, the diffusion activation energy should be reduced to the energy of interstitial migration.

Considering the effective activation energy (see Table 4), one can divide SPPs into two groups: Morelon-03, Yamada-00, Basak-03, Yakub-09 and MOX-07 have lower energy values 1.0–1.2 eV, while Walker-81, Busker-02, Nekrasov-08, Arima-05 and Goel-08 have higher values 1.7–2.5 eV. In this work we did not calculated migration energy values using the lattice statics method. However, for some potentials the values from previous works [37] [20] can be used for comparison. Their interstitial migration energies are close to our activation energies for the first group of potentials, while for Walker-81 and Busker-02 the lattice statics method underestimates the interstitial migration energy, probably due to its inherent difficulty of simulation of non-equilibrium states.

The superionic transition is often called premelting [21] due to a resemblance of anionic sublattice state to a liquid. Figs. 5 and 7 support this idea, since for MOX-07 the melting transition insignificantly changes rate and activation energy of oxygen diffusion. With other potentials DC also jumps at melting by no more than 50%. From Tables 4 and 5 it can be seen that the diffusion activation energy is halved after the melting for all five potentials of the second group. These potentials are also characterized by high density jump at melting (see [28]). Moreover, in Fig. 11 one can see that correlation of these two quantities is quite strong. Finally, for abnormal Yamada-00 SPP the activation energy is even slightly increased at melting.

## 4. Conclusion

In this work we carried out simulation of oxygen self-diffusion in three phase states of uranium dioxide, namely crystalline, superionic and melt. Dependence of diffusion coefficient (DC) on set of pair potentials (SPP), system size and boundary conditions was measured in a wide range of temperatures, which became possible due to our high-speed parallel implementation of molecular dynamics (MD) using graphics processors (GPU) [35] [36]. For the first time, low-temperature region (T < 1500 K) of the known experimental data [8] [9] was reached, which made possible direct comparison (i.e., without extrapolations). Observation of self-diffusion with DC down to $10^{-12}$ cm$^2$/s required simulation times up to 1000 ns (200 million MD steps), while the predecessors were bounded by $10^{-8}$ cm$^2$/s and 40 ps (see Table 1), correspondingly. Accurate tracking of the change in diffusion mechanisms (in particular, during the superionic transition) demanded a temperature step of 1 K, which was two orders of magnitude smaller than in most of the previous works. Such small step allowed calculating temperature dependence of diffusion activation energy $E_D(T)$ as derivative of Arrhenius plot (ln(D) vs. 1/kT), as well as the second derivative, which has an extreme.

Regardless of SPP the curve of DC has a form of hyperbola with two asymptotes for crystalline and superionic phases, which corresponds to S-shaped $E_D(T)$ curve with two plateaus for completely ordered and completely disordered anionic sublattice. Derivative of $E_D(T)$ has a form of λ-peak and its extreme has temperature equal to the temperature of λ-peak of isochoric heat capacity [1], which allows considering this temperature as the point of superionic transition. For MOX-07 and Yakub-09 potentials this temperature equals to 2600 K and 2700 K, correspondingly, which are close to the existing experimental data [48] [26].

MD simulations under periodic boundary conditions (PBC) allowed only three distinct mechanisms of oxygen self-diffusion, one for each of the phase states (i.e., crystalline, superionic and melt). These mechanisms were established by visual observation of ions movement and through analysis of $E_D(T)$ dependences.

Firstly, in periodically-translated crystals with ideal lattice (i.e., without surface or cavities) we have not found any long-lived intrinsic defects. So, the crystalline phase of this model, regardless of SPP and system size, is characterized by the exchange diffusion in the form of cyclic permutations of two or more anions with formation and recombination of instantaneous (short-term) anti-Frenkel defects. For all 10 SPPs the activation energy, ranging from 3.9 eV to 7.9 eV (see Tables 2 and 3), is close to the corresponding anti-Frenkel defect formation energy calculated by the lattice statics method. The best reproduction of experimental [44] and recommended [8] values of anti-Frenkel defect formation energy is given by MOX-07 and Morelon-03 SPPs.

In some earlier works, where the simulations were carried out under PBC without embedding artificial defects, the effective diffusion activation energy appeared to be close to the Matzke's recommendation of 2.6 eV [8]. However, interstitial and vacancy diffusion mechanisms proposed there [19] [21] were not directly registered but assumed from data approximation. Moreover, such low activation energies are in contradiction with the high anti-Frenkel defect formation energies calculated by the lattice statics method for the corresponding potentials, as in ideal lattice the diffusion cannot occur without defect formation. Results of the current work suggest that those low values were underestimated mainly due to coarse temperature step (100–250 K) and lack of simulations at sufficiently low temperatures (below 2200 K), since both factors hid the nonlinearity of the temperature dependences and the rising of statistical uncertainty with decreasing temperature. So, those assumptions [19] [21] about vacancy or interstitial mechanism of oxygen self-diffusion under PBC should be rejected.

However, these diffusion mechanisms occur in the presence of surface or artificial defects. In particular, MD simulations of isolated in vacuum nanocrystals with surface of energetically-optimal octahedral form using MOX-07 potentials made it possible to accurately reproduce the experimental dependences of DC (see Fig. 6). In this case, the curve of diffusion activation energy at temperatures below 2000 K shows deviation from the low-temperature plateau during the changeover of diffusion mechanism from exchange (with short-lived anti-Frenkel defects) to interstitial or vacancy (with long-lived defects). Artificial embedding of Schottky trivacancies into periodically-translated crystals also has led to similar changeover. This proves the need for surface or other source of defects in order to correctly simulate diffusion processes in the crystalline phase.

Secondly, at present there exist phenomenological hypotheses [26] [21] [47] of discontinuous transition between the two (crystalline and disordered) equilibrium states of anionic sublattice, which predict a narrow (20–50 K) λ-peak of the heat capacity with "infinite" height. In that case, a plot of oxygen DC in the Arrhenius coordinates would have a break corresponding to the stepwise change in the effective diffusion activation energy. In earlier works on diffusion simulation this superionic transition was not registered [18] [23] or demarcated unreliably [19] [20] [21] [22] [24]. On the contrary, temperature dependences of $E_D$ obtained in this work clearly have two plateaus of crystalline and superionic phases with wide (~1000 K) and gradual transition between them, which confirm smooth λ-peaks of heat capacity and linear thermal expansion coefficient obtained in our previous work [1]. The only exception was abnormal Yamada-00 SPP, which once again showed metastable coexistence of two phases with spontaneous step-wise changes of DC in the region of 2000–2600 K (see subfigure in Fig. 5). Thus, it is shown that in MD simulation of uranium dioxide the superionic phase transition has continuous nature (rather than first or second order) and occurs even in the absence of electronic and cationic defects (i.e., in the surfaceless crystals of rigid ions).

The superionic phase is characterized by saturated concentration of the intrinsic oxygen defects and low diffusion activation energy. This behavior is supported by the neutron scattering experiments [10], where the total fraction of defective anions also saturates to 20% near 2700 K. Correspondingly, in the superionic phase an interstitial diffusion of oxygen anions via "canals" formed by the cationic sublattice was observed. For five SPPs the diffusion activation energies are in range of 1–1.2 eV, which is close to the experimental estimates [8] [9] of the interstitial migration energy and the values calculated by the lattice statics method [37] [20], while the other five SPPs gave greater activation energies in range of 1.7–2.5 eV (see Table 4).

Finally, in the melt there was free movement of anions with minimal activation energy of 0.7–1.4 eV (see Table 5). Besides, the melting was accompanied by only minor jump in diffusion coefficients, which support the "premelting" hypothesis about high degree of oxygen sublattice disordering in the superionic phase [21].

At last, it is important to note that in the technologically significant temperature region of $UO_2$ crystalline phase the diffusion coefficients for various SPPs differ by orders of magnitude and not tens of percents like thermophysical characteristics [1] [28]. In this work the oxygen DC behavior closest to the experimental data is obtained for MOX-07 and Morelon-03 potentials. However, Morelon-03 SPP has computationally more complex form and gave worse reproduction of the thermophysical properties (see [1]). Besides, it was parameterized to reproduce the diffusion properties of $UO_2$, so independent confirmation of experimental data is given only by MOX-07 SPP. The other sets of pair potentials give the DC values by at least two orders of magnitude lower.

Table 1. Comparison of MD simulations of oxygen self-diffusion in $UO_2$.

| Source | Range of DC, $cm^2/s$ | Range of T, K | Step of T, K | Time per point, ps | Ion count |
|---|---|---|---|---|---|
| **This work (PBC)** | $4\times10^{-12} - 10^{-3}$ | 1400–7000 | 1–5 | $10-10^6$ | 1500 |
| **This work (IBC)** | $10^{-10} - 10^{-4}$ | 1400–3000 | 1–5 | $10-10^6$ | 4116 ** |
| Devynck 2012 [25] | $10^{-7} - 10^{-4}$ | 2000–2500 | 100 | $10-10^3$ | 1500 |
| Arima 2010 [32] | $10^{-8} - 10^{-4}$ | 2000–4700 | 250 | – | 1500 |
| Govers 2009 [23] | $10^{-7} - 10^{-4}$ | 2000–3000 | 250 | 25 | 20 736 |
| Goel 2008 [31] | $10^{-7} - 10^{-4}$ | 2100–3300 | 200 | 20 | 1500 |
| Kupryazhkin 2008 [22] | $10^{-7} - 10^{-4}$ | 2300–3600 | 100 | – | 4116 ** |
| Potashnikov 2007 [20] | $6\times10^{-10} - 10^{-3}$ | 1400–3900 | 50 | $10-10^5$ | 12 000 ** |
| Yakub 2007 [21] | $6\times10^{-9} - 10^{-3}$ | 2000–6000 | 100–200 | 10–40 | 2592 |
| Kupryazhkin 2004 [19] | $4\times10^{-8} - 10^{-4}$ | 1800–3600 | 150–250 | – | 1500 |
| Morelon 2003 [18] | $10^{-6} - 10^{-4}$ | 2200–3050 | 100 | 4 | 324 |

** – nanocrystal isolated in vacuum (i.e. isolated boundary conditions).

Table 2. Point defects formation energies calculated by the lattice statics method in comparison with diffusion activation energies ($E_D$) obtained from our MD simulations.

| SPP | SD, eV | FD, eV | AFD, eV | $E_D$, eV |
|---|---|---|---|---|
| Walker-81 | 7.8 (8.3) | 22.1 (22.3) | 5.9 (6.0) | 4.9 |
| Busker-02 | 14.7 | 29.0 | 8.4 | 7.9 |
| Nekrasov-08 | 8.7 | 21.0 | 5.7 | 5.2 |
| Morelon-03 | **8.0 (8.0)** | **15.6 (15.7)** | **3.9 (3.9)** | **3.9** |
| Yamada-00 | 12.9 (13.5) | 18.3 (18.5) | 5.8 (6.0) | 5.4 |
| Basak-03 | 10.3 (10.8) | 16.8 (17.0) | 5.8 (6.0) | 5.3 |
| Arima-05 | 14.5 | 23.0 | 7.9 | 7.7 |
| Goel-08 | 7.7 | 17.6 | 5.2 | 4.6 |
| Yakub-09 | 10.9 | 15.9 | 5.6 | 5.7 |
| MOX-07 | **9.8** | **15.6** | **4.1** | **3.9** |
| Read-11 | 12.4 | 25.2 (17.2) | 6.6 (4.6) | 6.3 |
| rec Matzke-87 [8] | 6.5±0.5 | 9.5±0.5 | 3.5±0.5 | – |
| DFT Nerikar-09 [39] | 7.6 | 15.1 | 3.95 | – |
| DFT Dorado-10 [40] | 10.6 | 14.6 | 6.5 | – |

( ) – values in parentheses are cited from [37] and [38] for comparison.
rec – recommendations based on experimental data.
DFT – results of *ab initio* calculations using Density Functional Theory.

Table 3. Characteristics of oxygen self-diffusion for crystalline phase of $UO_2$.

| SPP | T, K | $E_D$, eV | $D_0$, cm²/s |
|---|---|---|---|
| Walker-81 | 1900–2750 | 4.92±0.05 | 2.28E+02 |
| Busker-02 | 2800–3900 | 7.85±0.08 | 3.81E+03 |
| Nekrasov-08 | 1900–2550 | 5.20±0.12 | 1.54E+03 |
| Morelon-03 | 1500–2100 | 3.91±0.03 | 9.68E+02 |
| Yamada-00 | 1700–2000 | 5.40±0.17 | 5.77E+05 |
| Basak-03 | 1700–2350 | 5.27±0.04 | 9.41E+04 |
| Arima-05 | 2300–3100 | 7.65±0.06 | 1.31E+06 |
| Goel-08 | 1600–2100 | 4.59±0.08 | 5.25E+03 |
| Yakub-09 | 1800–2300 | 5.68±0.05 | 6.08E+05 |
| MOX-07 | 1400–2150 | 3.88±0.02 | 1.42E+03 |

Table 4. Characteristics of oxygen self-diffusion for superionic phase of $UO_2$.

| SPP | T, K | $E_D$, eV | $D_0$, cm²/s |
|---|---|---|---|
| Walker-81 | 4500–5000 | 2.20±0.02 | 1.76E–02 |
| Busker-02 | 6300–7100 | 2.53±0.02 | 1.02E–02 |
| Nekrasov-08 | 4500–5050 | 1.95±0.01 | 1.21E–02 |
| Morelon-03 | 3550–4250 | 1.15±0.01 | 4.52E–03 |
| Yamada-00 | 2200–5000 | 1.01±0.01 | 1.30E–03 |
| Basak-03 | 3400–4200 | 1.23±0.01 | 2.72E–03 |
| Arima-05 | 4050–4550 | 2.12±0.02 | 1.90E–02 |
| Goel-08 | 3300–3850 | 1.75±0.01 | 1.69E–02 |
| Yakub-09 | 3150–3750 | 1.22±0.01 | 3.80E–03 |
| MOX-07 | 3200–4000 | 0.97±0.01 | 3.14E–03 |

Table 5. Characteristics of oxygen self-diffusion for melted $UO_2$.

| SPP | T, K | $E_D$, eV | $D_0$, cm²/s |
|---|---|---|---|
| Walker-81 | 5000–6000 | 1.12±0.01 | 1.67E–03 |
| Busker-02 | 7100–8100 | 1.39±0.02 | 2.11E–03 |
| Nekrasov-08 | 5050–6050 | 1.01±0.01 | 1.67E–03 |
| Morelon-03 | 4250–5250 | 0.71±0.01 | 1.62E–03 |
| Yamada-00 | 5000–6000 | 1.20±0.01 | 2.39E–03 |
| Basak-03 | 4200–5200 | 1.03±0.01 | 2.20E-03 |
| Arima-05 | 4550–5550 | 1.08±0.01 | 2.06E–03 |
| Goel-08 | 3850–4850 | 0.86±0.01 | 1.65E–03 |
| Yakub-09 | 3750–4750 | 0.94±0.01 | 2.29E–03 |
| MOX-07 | 4000–5000 | 0.73±0.01 | 1.85E–03 |

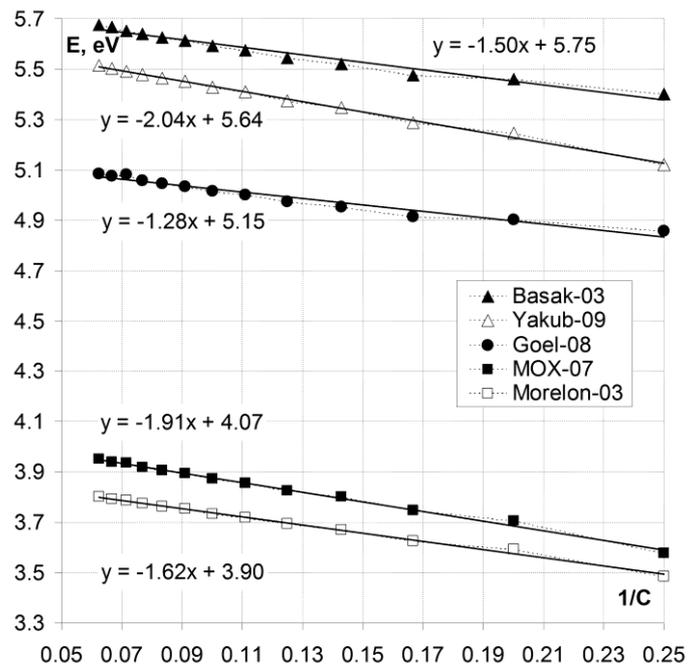

Fig. 1. Dependence of anti-Frenkel defect formation energy on reciprocal size of supercell.

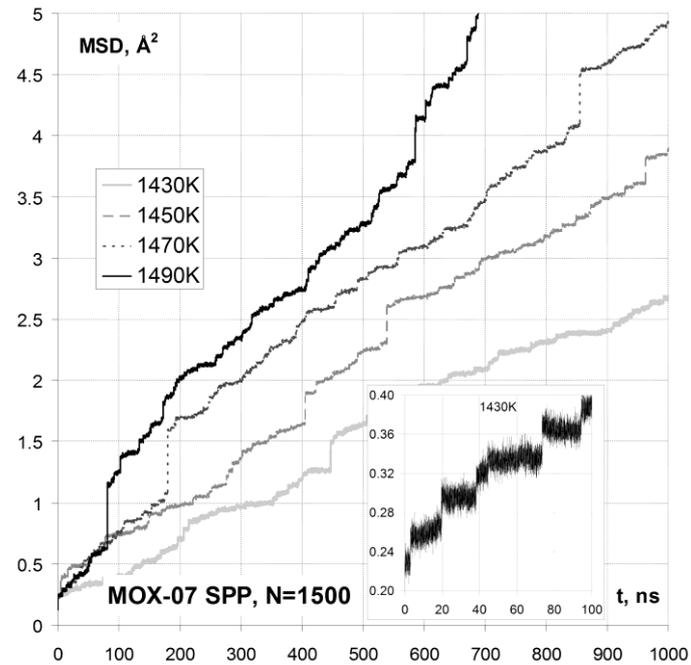

Fig. 2. Step-wise dependences of anion mean square displacement at sufficiently low temperatures (the subfigure shows close-up of clear staircase in 100 ns time interval).

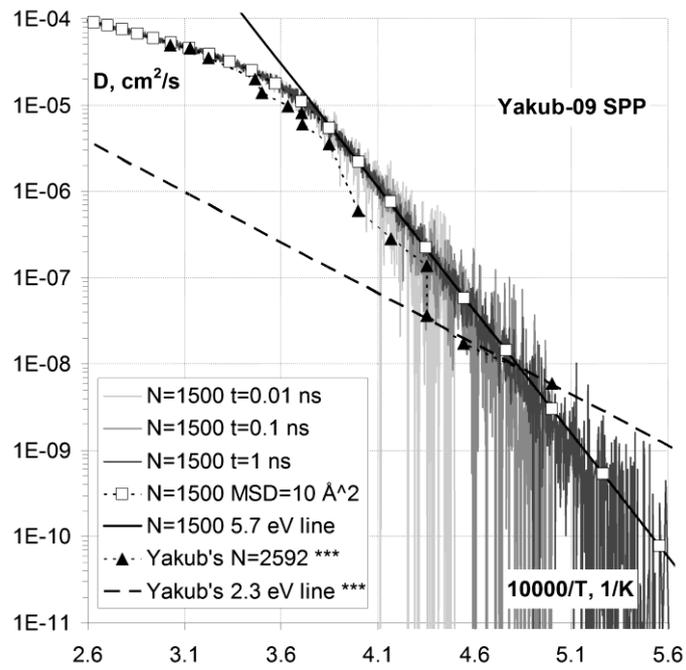

Fig. 3. Effect of simulation time on temperature dependence of anion self-diffusion coefficient (*** – data obtained with fixed simulation time in the work [21] are drawn for comparison).

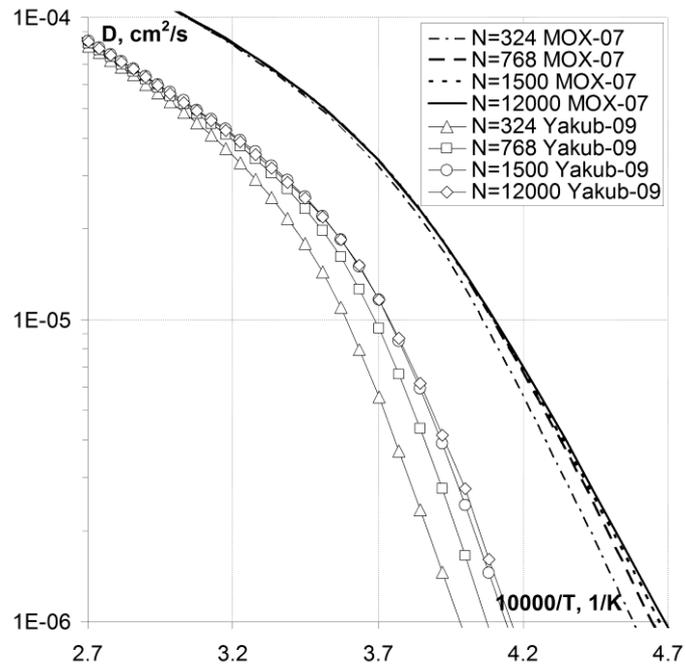

Fig. 4. Influence of supercell size on temperature dependence of anion self-diffusion coefficient.

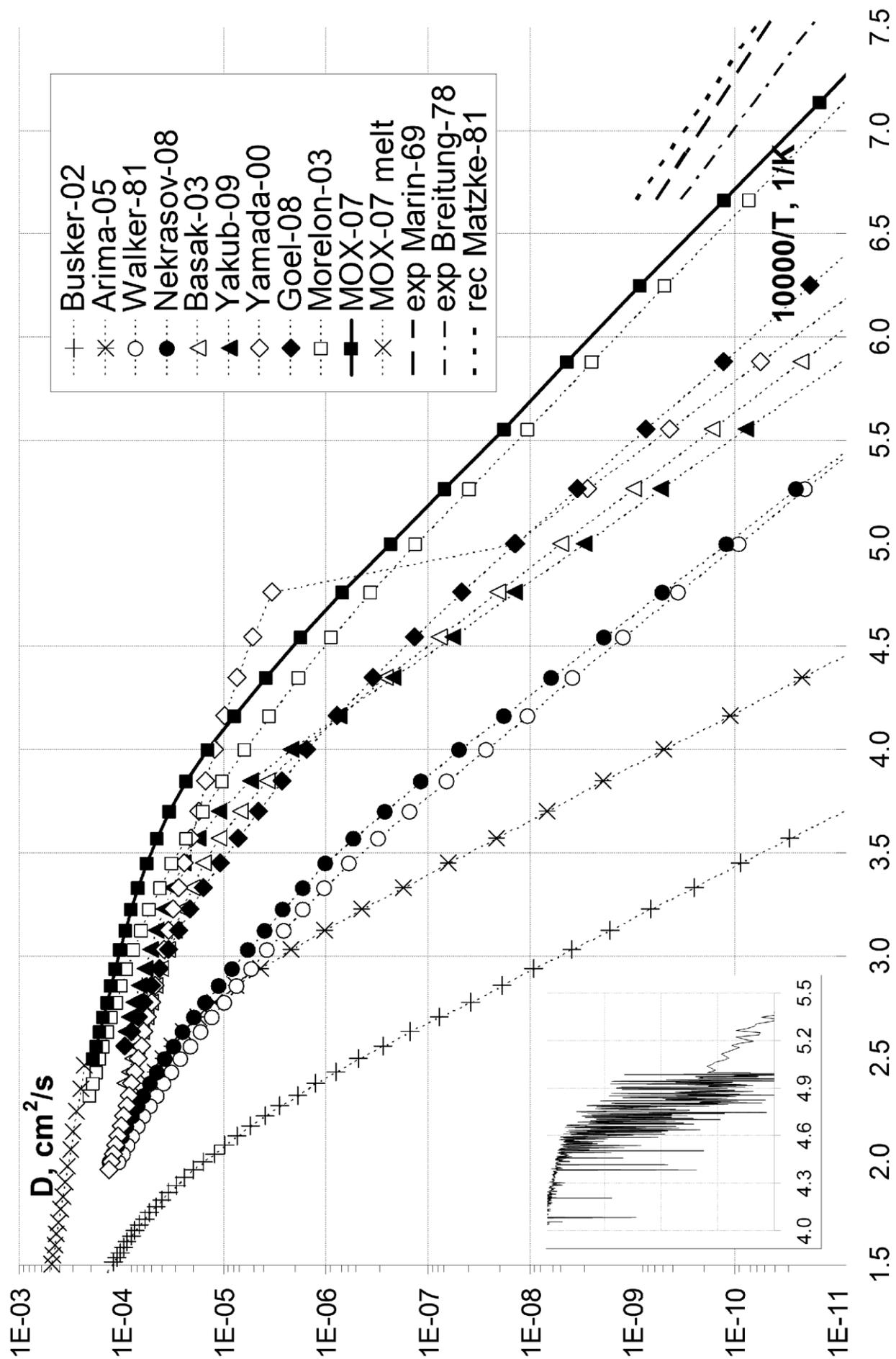

Fig. 5. Comparison of anion self-diffusion coefficients for different SPPs (subfigure with Yamada-00 SPP shows its anionic sublattice instability and abrupt phase transitions).

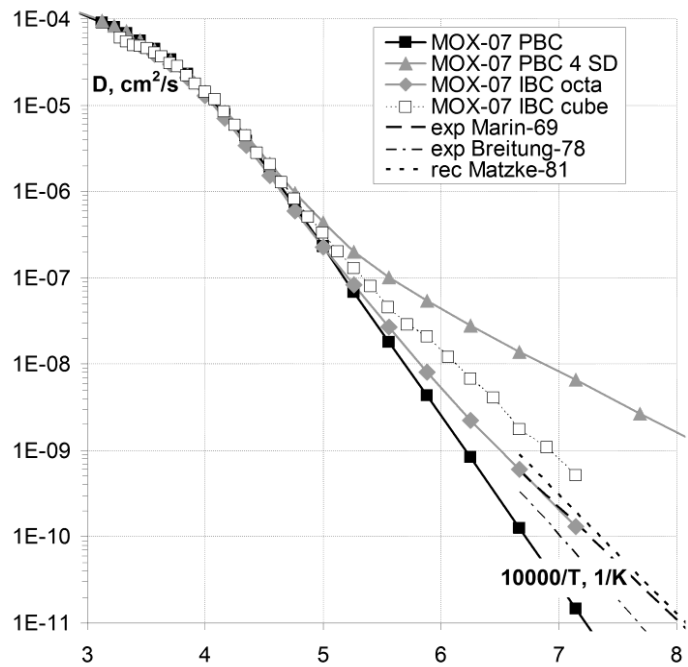

Fig. 6. Influence of surface and boundary conditions on qualitative diffusion behavior. The curve "PBC 4 SD" is obtained for supercell with cavity of 4 Schottky Defects bound together, "IBC octa" – for isolated nanocrystal of octahedral form and "IBC cube" – for isolated nanocrystal of cubic form (from [20]).

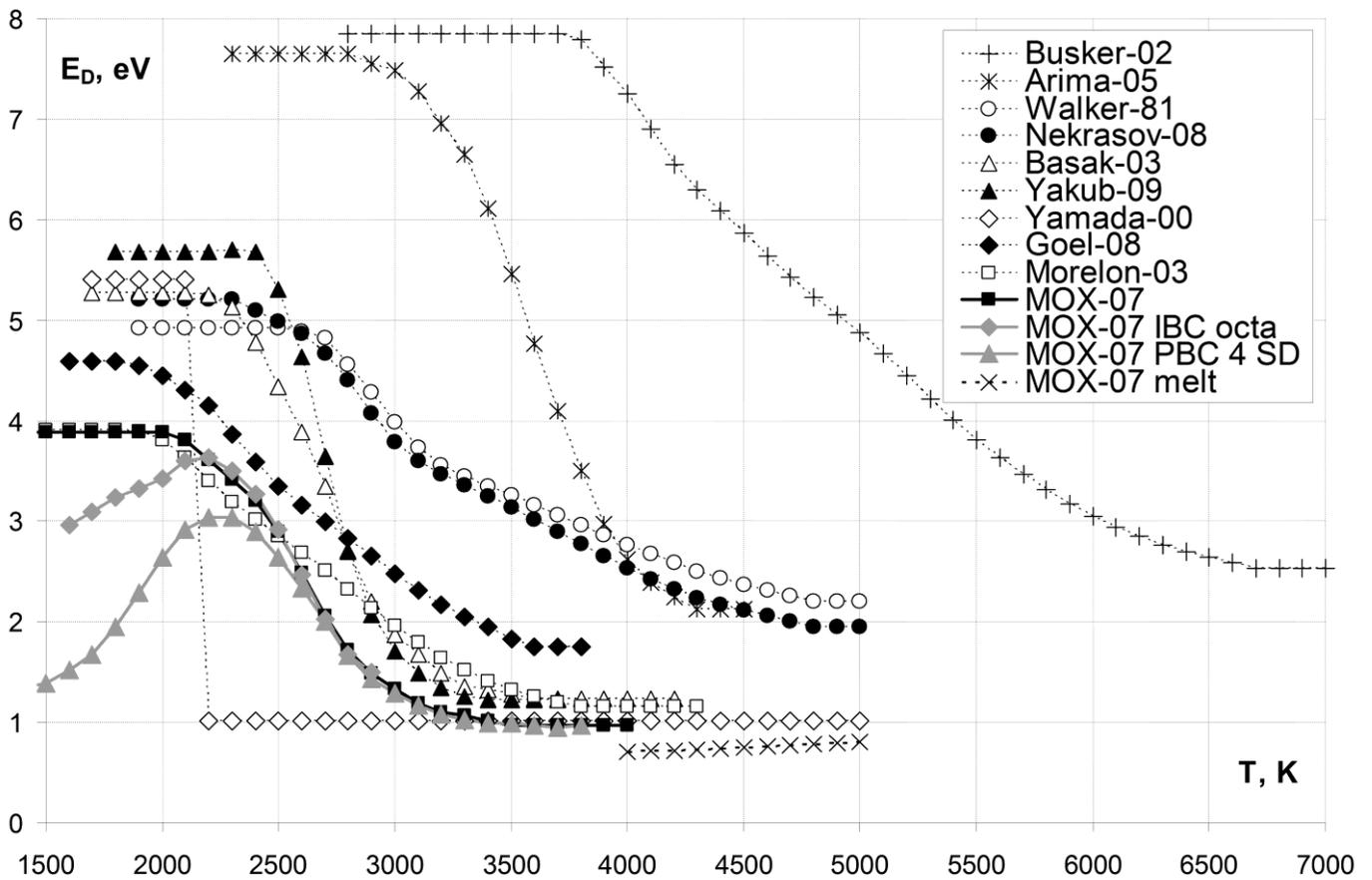

Fig. 7. S-shaped temperature dependence of diffusion activation energy.

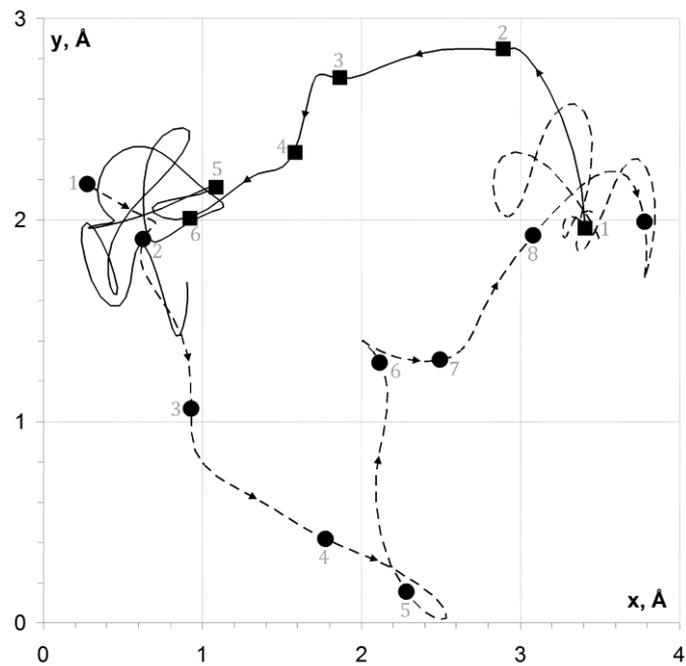

Fig. 8. Trajectories of two ions during sample exchange (one symbol for each ten MD steps).

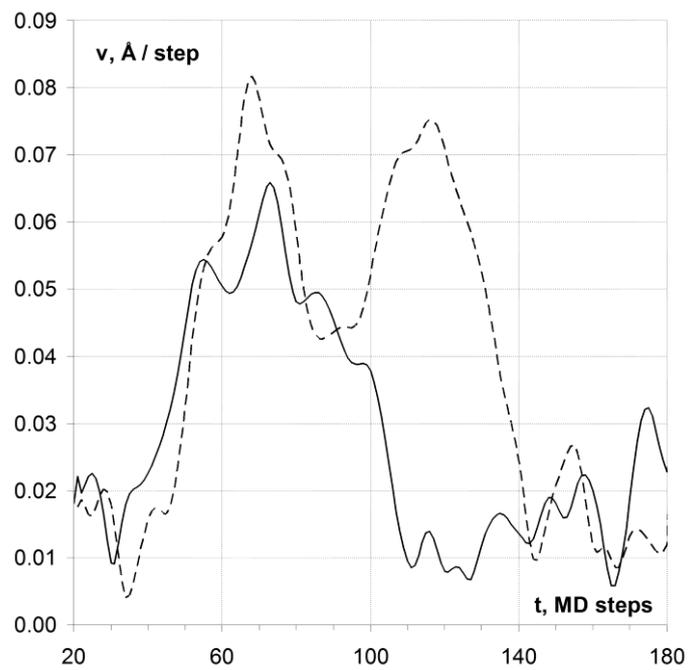

Fig. 9. Velocities of two ions during sample exchange.

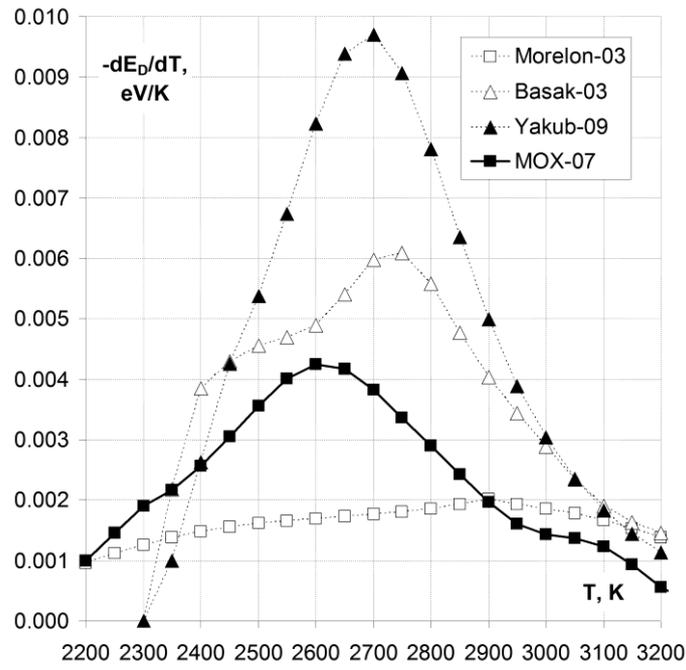

Fig. 10. λ-peaks on derivative of temperature dependence of diffusion activation energy.

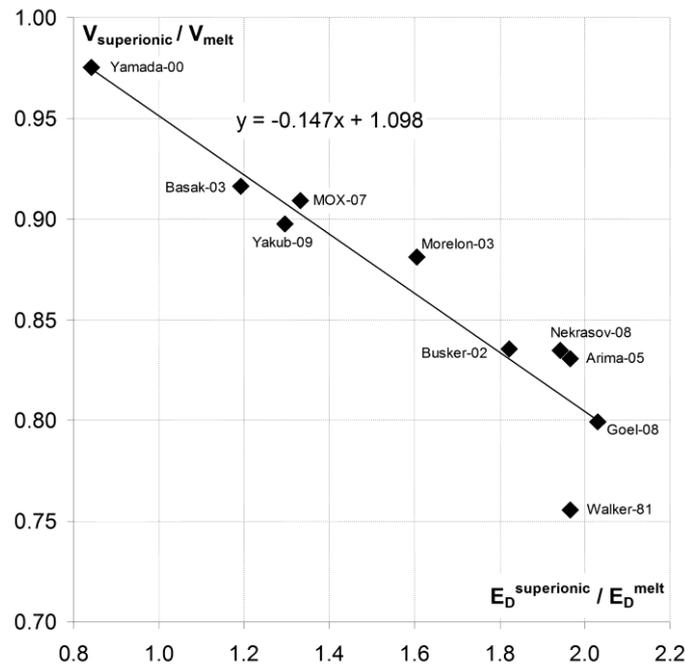

Fig. 11. Linear correlation of density jump and effective activation energy jump at melting.